\documentstyle[12pt]{article}
\def\be{\begin{eqnarray}}
\def\ee{\end{eqnarray}}
\def\ba{\begin{array}}
\def\ea{\end{array}}
\def\p{\phi}
\def\vp{\varphi}
\def\a{\alpha}

\def\pa{\partial}
\def\t{\tilde}

\def\G{{\cal G}}

\def\A{{\cal A}}
\def\X{{\cal X}}

\def\D{^{(D)}}
\begin{document}
\begin{center}
{\Large \bf {
Multi--dimensional IWP Solutions for\\
\vskip 3mm
Heterotic String Theory}}
\end{center}
\vskip 1cm
\begin{center}
{\bf \large  {Alfredo Herrera-Aguilar}}
\end{center}
\begin{center}
Joint Institute for Nuclear Research,\\
Dubna, Moscow Region 141980, Russia.\\
e-mail: herrera@thsun1.jinr.ru
\end{center}
\vskip 0.5cm
\begin{center}
and
\end{center}
\vskip 0.5cm
\begin{center}
{\bf \large  {Oleg Kechkin}}
\end{center}
\begin{center}
Institute of Nuclear Physics,\\
M.V. Lomonosov Moscow State University, \\
Moscow 119899, Russia, \\
e-mail: kechkin@monet.npi.msu.su
\end{center}
\begin{abstract}
We present extremal stationary solutions that generalize the
Israel--Wilson--Perj\'es class for the $d+3$--dimensional low--energy limit
of heterotic string theory with $n\geq d+1$ $U(1)$ gauge fields compactified
on  a $d$--torus. A rotating axisymmetric dyonic solution is obtained using
the matrix Ernst potential formulation and expressed in terms of a single
$d+1\times d+1$--matrix harmonic function.
By studying the asymptotic behaviour of the
field configurations we define the physical charges of the field system. The
extremality condition makes the charges to saturate the
Bogomol'nyi--Prasad--Sommerfield (BPS) bound. The gyromagnetic ratios of the
corresponding field configurations appear to have arbitrary values.
A subclass of rotating dyonic black
hole--type solutions arises when the NUT charges are set to zero. In the
particular case $d=1$, $n=6$, which correspond to $N=4$, $D=4$ supergravity,
the found dyon reproduces the supersymmetric dyonic solution constructed by
Bergshoeff {\it et al.}

\end{abstract}
\newpage
\section{Introduction}
In effective low energy theories of gravity derived from superstring theory
Einstein gravity is supplemented by additional fields such as the
Kalb--Ramond, gauge fields, and the scalar dilaton  which couples in a
non--trivial way to other fields
\cite{kir}
.
One of these theories is the bosonic sector of the heterotic string. This
model, when compactified from $D=d+3$ dimensions on a $d$--torus, can
be parametrized by the $(d+1)\times(d+1)$ and $(d+1)\times n$ Matrix Ernst
Potentials (MEP) $\X$ and $\A$ \cite{hk1}--\cite{hk2}
,
where $n$ is the number of Abelian vector fields. In order to have a
self--consistent quantum theory we must set $D=10$ and $n=16$,
but in this letter we shall leave these parameters arbitrary for the sake of
generality. However, there is one condition to be satisfied by them in order
to have a solution (see \cite{hk2}
for details), namely, $n\geq d+1$. The critical case is well described in the
framework of this formalism as well as the $N=4$, $D=4$ supergravity case,
when $d=1$ and $n=6$. Thus, the number of gauge fields is bounded below.

This letter is organized as follows. In Sec. 2 we present the effective
action of the low energy limit of heterotic string in terms of the MEP. This
fact allows one to map this action onto the stationary Einstein--Maxwell (EM)
action and apply classical procedures, commonly used with the EM theory, to
the heterotic string theory; in the framework of this approach we obtain in
Sec. 3 a stationary class of rotating dyonic solutions that generalizes
the Israel--Wilson--Perj\'es (IWP) class of the EM theory \cite{iwp}
by considering a linear dependence between the asymptotically flat potentials
$\X$ and $\A$. Then we define the physical charges of the field system by
studying the asymptotical behaviour of the $3$--fields. Furthermore, we
show that the physical charges of the obtained solutions saturate the BPS
bound as a consequence of the extremality condition. Among them we identify
axisymmetric rotating dyonic black hole--type solutions, endowed with
rotating axion, dilaton and Kalb--Ramond fields, when the NUT
charges vanish. In Sec. 4 we consider the particular case $N=4$,
$D=4$ supergravity and show that our solutions reproduce the
supersymmetric dyonic solutions of
\cite{bko}
.
Sec. 5 contains some concluding remarks and a brief discussion.

\section{Matrix Ernst Potentials}
The effective action of low energy limit of heterotic string theory is
\be
S\D\!=\!\int\!d\D\!x\!\mid\!
G\D\!\mid^{\frac{1}{2}}\!e^{-\p\D}\!(R\D\!+\!
\p\D_{;M}\!\p^{(D);M}
\!-\!\frac{1}{12}\!H\D_{MNP}H^{(D)MNP}\!-\!
\frac{1}{4}F^{(D)I}_{MN}\!F^{(D)IMN}),
\ee
where
\be
F^{(D)I}_{MN}\!=\!\pa_MA^{(D)I}_N\!-\!\pa _NA^{(D)I}_M, \quad
H\D_{MNP}\!=\!\pa_MB\D_{NP}\!-\!\frac{1}{2}A^{(D)I}_M\,F^{(D)I}_{NP}\!+\!
\mbox{\rm cycl. perms. of M,N,P.}
\nonumber
\ee
Here $G\D_{MN}$ is the multidimensional metric,
$B\D_{MN}$ is the anti--symmetric Kalb-Ramond field, $\p\D$ is the dilaton
and $A^{(D)I}_M$ denotes a set of $U(1)$ gauge fields ($I=1,\,2,\,...,n$).

After the Kaluza-Klein compactification on a $d$--torus, one obtains the
following set of three--dimensional fields
\cite{ms}-\cite{s1}
:

a) scalar fields
\be
G\!=\!(G_{pq}\!\equiv\!G\D_{p+3,q+3}),\quad
B\!=\!(B_{pq}\!\equiv\!B\D_{p+3,q+3}),\quad
A\!=\!(A^I_p\!\equiv\!A^{(D)I}_{p+3}),\quad
\p\!=\!\p\D\!-\!\frac{1}{2}{\rm ln\,|det}\,G|,
\ee
where the subscripts $p,q=1,2,...,d$.

b)tensor fields
\be
g_{\mu\nu}\!=\!e^{-2\p}\left(G\D_{\mu\nu}\!-\!G\D_{p+3,\mu}G\D_{q+3,\nu}G^{pq}
\right),\,\,\,
B_{\mu\nu}\!=\!B\D_{\mu\nu}\!-\!4B_{pq}A^p_{\mu}A^q_{\nu}\!-\!
2\left(A^p_{\mu}A^{p+d}_{\nu}-A^p_{\nu}A^{p+d}_{\mu}\right),
\nonumber
\ee
(we shall consider the ansatz when $B_{\mu\nu}=0$ in view of its
non--dynamical properties).

c)vector fields $A^{(a)}_{\mu}=
\left((A_1)^p_{\mu},(A_2)^{p+d}_{\mu},(A_3)^{2d+I}_{\mu}\right)$
($a=1,2,...,d,d+1,...,2d,2d+I$)
\be
(A_1)^p_{\mu}\!=\!\frac{1}{2}G^{pq}G\D_{q+3,\mu},\quad
(A_3)^{I+2d}_{\mu}\!=\!-\frac{1}{2}A^{(D)I}_{\mu}\!+\!A^I_qA^q_{\mu},\quad
(A_2)^{p+d}_{\mu}\!=\!\frac{1}{2}B\D_{p+3,\mu}\!-\!B_{pq}A^q_{\mu}\!+\!
\frac{1}{2}A^I_{p}A^{I+2d}_{\mu},
\nonumber
\ee
which can be dualized on-shell as follows
\begin{eqnarray}
\nabla\times\overrightarrow{A_1}&=&\frac{1}{2}e^{2\p}G^{-1}
\left(\nabla u+(B+\frac{1}{2}AA^T)\nabla v+A\nabla s\right),
\nonumber                          \\
\nabla\times\overrightarrow{A_3}&=&\frac{1}{2}e^{2\p}
(\nabla s+A^T\nabla v)+A^T\nabla\times\overrightarrow{A_1},
\nonumber                            \\
\nabla\times\overrightarrow{A_2}&=&\frac{1}{2}e^{2\p}G\nabla v-
(B+\frac{1}{2}AA^T)\nabla\times\overrightarrow{A_1}+
A\nabla\times\overrightarrow{A_3}.
\end{eqnarray}
The columns $u$ and $v$ have dimension $d$ with conponents $u_1$, $u_r$ and
$v_1$, $v_r$ ($r=2,3,...,d$), respectively, whereas the dimension of the
column $s$ is $n$. Thus, the final system is defined by the field variables
$G$, $B$, $A$, $\p$, $u$, $v$ and $s$.

At this stage it is convenient to introduce the matrix Ernst potentials
\be
\X=
\left(
\ba{cc}
-e^{-2\p}+v^TXv+v^TAs+\frac{1}{2}s^Ts&v^TX-u^T \cr
Xv+u+As&X
\ea
\right), \quad
\A=\left(
\ba{c}
s^T+v^TA \cr
A
\ea
\right),
\ee
where the $d \times d$ matrix potential $X=G+B+\frac{1}{2}AA^T$.
This pair of potentials allows us to express the
$3$-dimensional action in a quasi--EM form
\cite {m}
:
\be
S^{(3)}\!=\!\int\!d^3x\!\mid g\mid^{\frac{1}{2}}\!\{\!-\!R\!+\!{\rm Tr}\,[
\frac{1}{4}\left(\nabla \X\!-\!\nabla \A\A^T\right)\!\G^{-1}
\!\left(\nabla \X^T\!-\!\A\nabla \A^T\right)\!\G^{-1}
\!+\!\frac{1}{2}\nabla \A^T\G^{-1}\nabla \A]\},
\ee
where $\G=\frac{1}{2}\left(\X+\X^T-\A\A^T\right)$. The equations of motion
of the matter part of this action have the Ernst form \cite{e}
:
\be
&&\nabla^2\X -2(\nabla \X-\nabla \A \A^T)(\X+\X^T-\A\A^T)^{-1}\nabla \X=0,
\nonumber
\\
&&\nabla^2\A -2(\nabla \X-\nabla \A \A^T)(\X+\X^T-\A\A^T)^{-1}\nabla \A=0.
\ee
\section{Dyonic Rotating Black Hole--type Solutions}
In this Sec. we obtain a class of extremal solutions for the
equations of motion (6) which generalize the IWP class of the EM theory
following the procedure indicated in
\cite{hk2}.
We consider a linear dependence between the potentials $\A$ and $\X$, and
require them to be asymptotically flat, i.e.,
$\X_{\infty} \rightarrow \Sigma$ and $\A_{\infty} \rightarrow 0$,
where $\Sigma=diag(-1,-1,1,1,...,1)$. Thus, the matrix Ernst potentials are
related by
\be
\A=(\Sigma - \X)b,
\ee
where $b$ is an arbitrary constant $d+1\times n$--matrix.
By substituting (7) into the action
(5) and setting the Lagrangian of the system to zero (it implies that
$R_{ij}=0$), we get the following condition to be satisfied
\be
bb^T=-\Sigma /2.
\ee
Indeed, {\it both} equations of motion (6) reduce to the Laplace equation
in Euclidean $3$--space
\be
\nabla ^2 [(\Sigma + \X)^{-1}]=0.
\ee
The solutions for this equation are well known and in the simplest case one
can consider the harmonic function
\be
\frac{2}{\Sigma + \X}= \Sigma +Re\frac{M}{R},\quad {\rm where}\quad
R^2=x^2+y^2+(z+i\a)^2,
\ee
$M$ is a complex $d+1$--dimensional constant matrix with arbitrary
components $m_{\t p,\t q}=\t m_{\t p,\t q}+i \t n_{\t p,\t q}$,
and $\a$ is a real constant. We choose $M$ and $R$ in this way
in order to deal with rotating black hole--type solutions (in this case we
have a ring singularity) when the NUT charges of the field system are set
to zero. This is in contrast with the results obtained in
\cite{hk2} and \cite{hk3}
,
where the four-- and five--dimensional stationary classes of solutions become
static when the NUT charges vanish. In a forthcoming paper we will
investigate solutions where both $M$ and  $b$ have a more general form. Such
an ansatz leads to a richer class of solutions for the theory under
consideration.

In order to obtain a real value of the potential $\A$ (see Eqs. (7) and (8))
we can require just the first two rows of $b$ to
be real (leaving the remaining rows imaginary), then we perform the matrix
product (7) and set the factors that multiply the imaginary components of $b$
to zero. It turns out that this condition imposes the following restriction
on the matrix $M$
\be
M=\left(
\ba{ccccc}
m_{11}    & m_{12}    & 0 & \cdots  & 0 \cr
m_{21}    & m_{22}    & 0 & \cdots  & 0 \cr
m_{r+1,1} & m_{r+1,2} &   & 0_{d-1} &
\ea
\right),
\ee
where $0_{d-1}$ denotes a $(d-1)$--dimensional square array of zeroes. It is
not difficult to check that this  procedure leads to real solutions for
the potential $\A$.

At this stage  one is able to calculate the $3$--fields
$G$, $B$, $A$, $\p$, $u$, $v$ and $s$. By studying their asymptotic behaviour
one can establish the following relation between the integration
constants and the physical parameters of the theory
\be
G\sim
\left(
\ba{cc}
-\left(1+\frac{2\t m_{22}}{R_{as}}\right) & \frac{\t m^T_{r+1,2}}{R_{as}}  \cr
\frac{\t m_{r+1,2}}{R_{as}} &  1_{d-1}
\ea
\right)=
\left(
\ba{cc}
-\left(1-\frac{2m}{R_{as}}\right) &  \frac{C^T_r}{R_{as}}  \cr
\frac{C_r}{R_{as}} &  1_{d-1}
\ea
\right),
\nonumber
\ee
\be
B\sim
\left(
\ba{cc}
0 & -\frac{\t m^T_{r+1,2}}{R_{as}}  \cr
\frac{\t m_{r+1,2}}{R_{as}} &  0_{d-1}
\ea
\right)=
\left(
\ba{cc}
0 &  -\frac{C^T_r}{R_{as}}  \cr
\frac{C_r}{R_{as}} &  0_{d-1}
\ea
\right), \quad \p\sim -\frac{\t m_{11}}{R_{as}}=\frac{D}{R_{as}},
\nonumber
\ee
\be
A=\left(
\ba{l}
A^{I}_t \cr
A^{I}_r
\ea
\right)
\sim
\left(
\ba{c}
2(\t m_{21}b_{1I}+\t m_{22}b_{2I})/R_{as} \cr
-2(\t m_{r+1,1}b_{1I}+\t m_{r+1,2}b_{2I})/R_{as}
\ea
\right)=
\left(
\ba{c}
Q^{I}_e/R_{as} \cr
Q^{I}_r/R_{as}
\ea
\right),
\nonumber
\ee
\be
u_1\sim \frac{\t m_{12}-\t m_{21}}{R_{as}}=\frac{N}{R_{as}}, \quad
v_1\sim \frac{\t m_{12}+\t m_{21}}{R_{as}}=\frac{C_1}{R_{as}},
\nonumber
\ee
\be
u_r=v_r\sim \frac{\t m_{r+1,1}}{R_{as}}=\frac{N_r}{R_{as}}, \quad
s^I\sim 2\frac{\t m_{11}b_{1I}+\t m_{12}b_{2I}}{R_{as}}=
\frac{Q^{I}_m}{R_{as}},
\ee
where $b_{iI}b_{jI}=\delta_{ij}/2$, $i,j=1,2$; $m$ is the ADM mass, $D$ is
the dilaton, $N$ and $N_r$ are $d$ NUT charges, $C_1$ is a scalar (axion)
charge, $C_r$ are $d-1$ Kaluza--Klein charges, $Q^{I}_e$ and $Q^{I}_m$ are
two sets of $n$ electric and magnetic charges, and $Q^{I}_r$ are $d-1$ sets
of $n$ charges that come from the extra dimensions of the electromagnetic
sector; and $R_{as}=\sqrt{x^2+y^2+z^2}$. The extremality character of the found solutions makes these charges
to saturate the BPS bound
\be
4(D^2+m^2)+2(N^2+C^2_1)+\sum^d_{r=2} (Q_r^{I})^2=
(Q_e^{I})^2+(Q_m^{I})^2+4\sum^d_{r=2}(N^2_r+C^2_r),
\ee
where a summation under $I$ is understood. This means that the
attractive forces are precisely balanced by the repulsive forces in the
field configuration.

In order to write down the explicit form of a single point--like
solution in terms of the multidimensional variables we must calculate all
vector $3$--fields using the dualization formulae (3). After some
algebraic manipulations we obtain
\be
2\nabla\times\overrightarrow A_1=Re\nabla
\left[\!
\left(\!
\ba{c}
m_{21}-m_{12} \cr
m_{r+1,1}
\ea\!
\right)\!
\frac{1}{R}
\right]+
\left(\!
\ba{c}
\sigma \cr
0_r
\ea\!
\right)\!Im
\left(
\frac{1}{\overline R}\nabla\frac{1}{R}\!
\right),
\nonumber
\ee
\be
2\nabla\times\overrightarrow A_2=Re\nabla
\left[\!
\left(\!
\ba{c}
-(m_{12}+m_{21}) \cr
m_{r+1,1}
\ea\!
\right)\!
\frac{1}{R}
\right]+
\left(\!
\ba{c}
\sigma \cr
0_r
\ea\!
\right)\!Im
\left(
\frac{1}{\overline R}\nabla\frac{1}{R}\!
\right),
\nonumber
\ee
\be
\nabla\times\overrightarrow A_{3}^I=
\left\{
Re\left(\nabla\frac{m_{11}}{R}\right)b_{1I}+
\left[Re\left(\nabla\frac{m_{12}}{R}\right)
-\sigma Im\left(\frac{1}{\overline R}\nabla\frac{1}{R}\right)
\right]b_{2I}\right\},
\ee
where $\sigma \!=\!\t m_{11}\t n_{12}\!+\!\t m_{21}\t n_{22}\!-\!
\t m_{12}\t n_{11}\!-\!\t m_{22}\t n_{21}\!+\!
\sum_r(\t m_{r+1,2}\t n_{r+1,1}\!-\!\t m_{r+1,1}\t n_{r+1,2})$;
$b_{1I}$ and $b_{2I}$ being the first two rows of the matrix $b$.

It is natural to write down the solutions in terms of oblate spheroidal
coordinates defined by
\be
x=\sqrt{\rho^2+\a^2}\sin\theta \cos\vp,\quad
y=\sqrt{\rho^2+\a^2}\sin\theta \sin\vp,\quad
z=\rho \cos\theta.
\ee
Then, the $3$--interval reads
\be
ds^2_{3}=(\rho^2+\a^2\cos^2\theta)(\rho^2+\a^2)^{-1}d\rho^2+
(\rho^2+\a^2\cos^2\theta)d\theta^2+
(\rho^2+\a^2)\sin^2\theta d\vp^2
\ee
and only the $A^{(a)}_{\vp}$ components do not vanish \footnote{In fact we
have imposed the axial symmetry with respect to $z$ and have chosen
$R=\rho+i\a\cos\theta$.}:
\be
2A_{1\vp}=(\rho^2\!+\!\a^2\cos^2\theta)^{-1}\!
\left\{\!
\left(\!
\ba{c}
\t m_{21}-\t m_{12} \cr
\t m_{r+1,1}
\ea\!
\right)\!
(\rho^2\!+\!\a^2)\cos\theta\!+\!
\left[\!
\left(\!
\ba{c}
\t n_{12}\!-\!\t n_{21} \cr
-\t n_{r+1,1}
\ea\!
\right)\!
\rho\!+\!\frac{1}{2}\!
\left(\!
\ba{c}
\sigma \cr
0_r
\ea\!
\right)\!
\right]\!
\a\sin^2\theta\!
\right\}\!,
\nonumber
\ee
\be
2A_{2\vp}=(\rho^2\!+\!\a^2\cos^2\theta)^{-1}\!
\left\{\!
\left(\!
\ba{c}
-(\t m_{12}+\t m_{21}) \cr
\t m_{r+1,1}
\ea\!
\right)\!
(\rho^2\!+\!\a^2)\cos\theta\!+\!
\left[\!
\left(\!
\ba{c}
\t n_{12}\!+\!\t n_{21} \cr
-\t n_{r+1,1}
\ea\!
\right)\!
\rho\!+\!\frac{1}{2}\!
\left(\!
\ba{c}
\sigma \cr
0_r
\ea\!
\right)\!
\right]\!
\a\sin^2\theta\!
\right\}\!,
\nonumber
\ee
\be
A_{3\vp}^I&=&(\rho^2+\a^2\cos^2\theta)^{-1}
\left\{
\left[\t m_{11}(\rho^2+\a^2)\cos\theta-
\t n_{11}\rho\a\sin^2\theta\!\right]b_{1I}+
\right.
\nonumber\\
&+&\left.
\left[\t m_{12}(\rho^2+\a^2)\cos\theta-
(\t n_{12}\rho+\sigma /2)\a\sin^2\theta\right]b_{2I}
\right\}.
\ee
From here we see that there exist $2d+n$ angular momenta defined by
\be
A^{(a)}_{\vp}\sim-\frac{\t n^{(a)}\a\sin^2\theta}{\rho}=
\frac{2J^{(a)}\sin^2\theta}{\rho},
\ee
where the parameters $\t n^{(a)}$ are arbitrary in the general case and hence,
the gyromagnetic ratios of the corresponding field configurations turn out to
be arbitrary as well in the context of this approach. In the general case the
expressions for the multidimensional fields depend on the arbitrary parameters
$\t n_{pq}$, as it takes place for the $A^{(a)}_{\vp}$ components. However, in
order to obtain the four--dimensional solutions constructed in \cite{bko}
(and their direct generalization to the multidimensional case) we set
\be
2\t n_{11}\!=\!(C_1\!-\!N),\,\,\, \t n_{12}\!=\!-m,\,\,\,
\t n_{21}\!=\!D,\,\,\, 2\t n_{22}\!=\!-(C_1\!+\!N),\,\,\,
\t n_{r+1,1}\!=\!C_r,\,\,\, \t n_{r+1,2}\!=\!-N_r,
\ee
and quote the full solution corresponding to this special case:
\be
ds^2=G_{MN}dx^{M}dx^{N}=G_{pq}\left(dx^{p+3}+\omega^{(p)}
d\vp\right) \left(dx^{q+3}+\omega^{(q)} d\vp\right)+
e^{2\p}g_{\mu\nu}dx^{\mu}dx^{\nu},
\ee
where the symmetric matrix $G_{pq}$ has the components
\be
G_{11}\!=\!-\Delta^{-2}
\left[(\rho^2\!+\!\a^2\cos^2\theta)\!+\!2D\rho\!+\!(N-C_1)\a\cos\theta\!+\!
D^2\!+\!(C_1-N)^2/4\right]\!(\rho^2+\a^2\cos^2\theta),
\nonumber
\ee
\be
G_{1r}\!=\!\Delta^{-2}\left\{(N_r\rho+C_r\a\cos\theta)
\left[(C_1\rho+(D-m)\a\cos\theta)\!+\!D(C_1+N)/2\!+\!m(C_1-N)/2\right]\right.+
\nonumber
\ee
\be
\left.(C_r\rho-N_r\a\cos\theta)
\left[(\rho^2+\a^2\cos^2\theta)+
2D\rho+(N-C_1)\a\cos\theta+D^2+(C_1-N)^2/4\right]
\right\},
\nonumber
\ee
\be
G_{rr'}=\delta_{r+1,r'+1}-\Delta^{-2}(\rho^2+\a^2\cos^2\theta)^{-2}
\left\{\left[(N_r\rho\!+\!C_r\a\cos\theta)
(\rho^2+\a^2\cos^2\theta)+\Delta_{2r}\right]\times\right.
\nonumber
\ee
\be
\left[(N_{r'}\rho\!+\!C_{r'}\a\cos\theta)(\rho^2+\a^2\cos^2\theta)\!+\!
\Delta_{2r'}\right]\!+\!
\left[(C_r\rho\!-\!N_r\a\cos\theta)(\rho^2\!+\!\a^2\cos^2\theta)\!+\!
\Delta_{1r}\right]\!\times
\nonumber
\ee
\be
\left.\left[(C_{r'}\rho-N_{r'}\a\cos\theta)(\rho^2+\a^2\cos^2\theta)+
\Delta_{1r'}\right]
\right\},
\nonumber
\ee
the conformal multiplier has the form
\be
e^{2\p}=1+\frac{2D\rho+(N-C_1)\a\cos\theta}{\rho^2+\a^2cos^2\theta}+
\frac{\delta_0}{(\rho^2+\a^2cos^2\theta)^2}
\ee
and the components of the rotational vector are defined by
\be
\omega^{(1)}=-(\rho^2+\a^2\cos^2\theta)^{-1}
\left\{N(\rho^2+\a^2)\cos\theta+
\left[(m+D)\rho-\sigma /2\right]\a\sin^2\theta\right\},
\nonumber
\ee
\be
\omega^{(r)}=(\rho^2+\a^2\cos^2\theta)^{-1}
\left[N_r(\rho^2+\a^2)\cos\theta-
C_r\a\rho\sin^2\theta\right],
\nonumber
\ee
where we have introduced the notations
$\Delta\!=\!\rho^2\!+\!\a^2cos^2\theta\!+\!(D\!+\!m)\rho\!+
\!N\a\cos\theta\!+\!Dm\!-\!(C^2_1\!-\!N^2)/4$,
$2\Delta_{1r}=(C_r\rho-N_r\a\cos\theta)(2D\rho+(N-C_1)\a\cos\theta)+
(N_r\rho+C_r\a\cos\theta)\left[(C_1+N)\rho-2m\a\cos\theta\right]$,
$2\Delta_{2r}=
(N_r\rho+C_r\a\cos\theta)(2m\rho+(C_1+N)\a\cos\theta)+
(C_r\rho-N_r\a\cos\theta)\left[(C_1-N)\rho+2D\a\cos\theta\right]$,
$\delta_0\!=\!\left[\!D^2\!+\!(C_1\!-\!N)^2/4\!\right]
\!(\rho^2\!+\!\a^2cos^2\theta)\!-\!(N_r\rho\!+\!C_r\a\cos\theta)^2$
and $\sigma =2mD+C_r^2+N_r^2+(N^2-C_1^2)/2$.

From Eq. (20) we see that the interval adopts the form of an axisymmetric
rotating black hole solution when the NUT charges vanish.

The only non--vanishing components of the multidimensional matter fields are
\be
B_{r1}=\Delta^{-1}(\rho^2+\a^2\cos^2\theta)^{-1}
\left[(C_r\rho-N_r\a\cos\theta)
(\rho^2+\a^2\cos^2\theta)+\Delta_{1r}\right],
\nonumber
\ee
\be
A^{I}_t\!=\!\Delta^{-1}\left\{
\left[(C_1\!-\!N)\rho\!+\!2D\a\cos\theta\right]
b_{1I}\!-\!\left[2m\rho\!+\!(C_1\!+\!N)\a\cos\theta\!+\!
2mD\!+\!(N^2\!-\!C_1^2)/2\right]b_{2I}\right\},
\nonumber
\ee
\be
A^{I}_r=-2\Delta^{-1}(\rho^2+\a^2\cos^2\theta)^{-1}\left\{
\left[N_r\rho+C_r\a\cos\theta)(\rho^2+\a^2\cos^2\theta)
+\Delta_{2r}\right]b_{1I}\right.+
\nonumber
\ee
\be
\left.\left[(C_r\rho-N_r\a\cos\theta)(\rho^2+\a^2\cos^2\theta)+
\Delta_{1r}\right]b_{2I}\right\},
\nonumber
\ee
\be
\p^{(D)}\!=\!{\rm ln}\!
\left\{\!\Delta^{-1}(\rho^2\!+\!\a^2\!\cos^2\theta)^{-1}\!
\left[\!(\rho^2\!+\!\a^2\!\cos^2\theta)^2\!+\!
\left(\!2D\rho\!+\!(N\!-\!C_1)\!\a\cos\theta\!\right)\!
(\rho^2\!+\!\a^2\!\cos^2\theta)\!+\!\delta_0\!\right]\!
\right\},
\nonumber
\ee
\be
B^{(D)}_{t\vp}=(\rho^2+\a^2\cos^2\theta)^{-1}
\left\{-A^{I}_t\left[Q^{I}_m(\rho^2+\a^2)\cos\theta-
\left(Q_e^{IT}\rho+\sigma b_{2I}^T/2\right)\a\sin^2\theta\right]/2\right.+
\nonumber
\ee
\be
\left.B^T_{r1}\left[N_r(\rho^2+\a^2)\cos\theta\!-\!
C_r\a\rho\sin^2\theta\right]
\!-\!\left[C_1(\rho^2+\a^2)\cos\theta\!+\!\left((m\!-\!D)\rho\!-\!\sigma /2\right)
\a\sin^2\theta\right]\right\},
\nonumber
\ee
\be
B^{(D)}_{r\vp}=(\rho^2+\a^2\cos^2\theta)^{-1}
\left\{-A^{I}_r\left[Q^{I}_m(\rho^2+\a^2)\cos\theta-
\left(Q_e^{IT}\rho+\sigma b_{2I}^T/2\right)\a\sin^2\theta\right]/2\right.-
\nonumber
\ee
\be
\left.B_{r1}\left[N(\rho^2+\a^2)\cos\theta-
\left((m+D)\rho+\sigma /2\right)\a\sin^2\theta\right]+
\left[N_r(\rho^2+\a^2)\cos\theta-
C_r\a\rho\sin^2\theta\right]\right\},
\nonumber
\ee
\be
A^{(D)I}_{\vp}=(\rho^2+\a^2\cos^2\theta)^{-1}
\left\{-\left[Q^{I}_m(\rho^2+\a^2)\cos\theta-
\left(Q_e^{IT}\rho+\sigma b_{2I}^T/2\right)\a\sin^2\theta\right]
\right.+A^{IT}_r\times
\nonumber
\ee
\be
\left.\!\left[N_r(\rho^2\!+\!\a^2)\cos\theta\!-\!
C_r\a\rho\sin^2\theta\right]\!\!-\!\!
A^{IT}_t\!\left[N(\rho^2\!+\!\a^2)\cos\theta\!-\!
\left((m+D)\rho+\sigma /2\right)\a\sin^2\theta\right]\right\},
\ee
where $b_{1I}=-\frac{1}{4\Delta_{12}}[(C_1+N)Q_e^{I}+2mQ_m^{IT}]$,
$b_{2I}=-\frac {1}{4\Delta_{12}}[2DQ_e^{I}+(C_1-N)Q_m^{IT}]$, and
$\Delta_{12}=mD-(C_1^2-N^2)/4$. One can see that the multidimensional
components of the fields $B^{(D)}_{t\vp}$, $B^{(D)}_{r\vp}$ and
$A^{(D)I}_{\vp}$ are non--trivial at spatial infinity as it takes place
for the magnetically charged configurations of the ordinary EM theory.

By counting the number of independent parameters which parametrize the
physical charges of the solution we see that matrix $M$ contributes with
$2(d+1)$ items in the framework of our ansatz. On the other hand, matrix
$b$ provides $2n-3$ independent parameters since only its first two rows
affect the solution (these rows are normalized and orthogonal to each other
in view of Eq. (8)). We have the rotational parameter $\a$ as well. Thus we
have in total $2(d+n)$ independent integration constants which define
charges of the field system.

Thus, our solution can be interpreted as an asymptotically Taub--NUT rotating
field configuration with axial symmetry formed by: the Einstein mass
$m_{E}\!=\!D\!+\!m$, the Kaluza--Klein charges $C_r$, their corresponding
NUT charges $N$ and $N_r$, the multi--dimensional dilaton
with charge $D^{(D)}\!=\!D\!-\!m$, the axion charge $C_1$,
the electromagnetic charges $Q_e^{I}$, $Q_m^{I}$ and $Q_r^{I}$ (the
fields with magnetic charges are non--trivial at spatial infinity as in
ordinary EM theory) and the antisymmetric Kalb--Ramond charges $C_r$ which
turn out to be equal to the Kaluza--Klein charges in our ansatz.  The total
number of independent charges is equal to $2(d+n)$ in the framework of the
ansatz under consideration.

\section{N=4, D=4 Supergravity}

In the particular case $d=1$, $n=6$ the considered action corresponds to the
bosonic sector of $N=4$, $D=4$ supergravity. A supersymmetric generalization
of the IWP solutions for such a theory was constructed by Bergshoeff
{\it et al.} \cite{bko}
choosing as ansatz two arbitrary complex harmonic functions.
In this Sec. we show that in the case of a single point--like source, our
solutions reproduce these solutions. In order to do so, it is convenient
to switch to the Einstein frame:
\be
ds_E^2=e^{-\p^{(4)}}ds_{str}^2=-\frac{(\rho^2+\a^2\cos^2\theta)}{\Delta}
\left(dt-\omega_{\vp}d\vp\right)^2+\frac{\Delta}{(\rho^2+\a^2\cos^2\theta)}
g_{\mu\nu}dx^{\mu}dx^{\nu},
\ee
where $e^{\p^{(4)}}\!=\!\Delta^{-1}\!\left[\rho^2\!+\!\a^2\!\cos^2\theta\!+
\!2D\rho\!+\!(N\!-\!C_1)\a\cos\theta\!+\!D^2\!+\!(C_1\!-\!N)^2/4\right]$
is the conformal factor and
$\omega_{\vp}\!=\!(\rho^2\!+\!\a^2\!\cos^2\theta)^{-1}\!\left\{\!
N(\rho^2\!+\!\a^2)\cos\theta\!+\!\left[m_{E}\rho\!-\!mD\!+\!
(C_1^2\!-\!N^2)/4\right]\a\sin^2\theta\right\}$ is the
angular velocity of our rotating object; from here we see that
$m_{E}/2$, $N/2$ and $(D^{(4)}+iC_1)/2$ are the mass, the NUT
charge and complex axion--dilaton charges of
\cite{bko}
, respectively.
The non--trivial components of the matter fields are given by the second,
fourth, fifth and seventh relations of Eq. (22) when the extra--dimensional
parameters vanish.

From Eq. (18) it is clear that the rotating axion charge generates the dipole
momentum
\be
J_a=D^{(4)}\a/4,
\ee
whereas the rotating $n$ electric charges induce a magnetic field and
originate the momenta
\be
J^{I}=-Q_e^{I}\a/4.
\ee
Thus, this particular solution corresponds to an asymptotically Taub--NUT
rotating field configuration with axial symmetry where the Einstein mass is
endowed with the NUT charge, the axion and dilaton charges and two sets of
$n$ electric and $n$ magnetic charges which rotate together with it.  The
fields generated by the charges $N$, $C_1$ and $Q_m^{I}$ do not vanish at
spatial infinity having a Dirac string peculiarity; they are the NUT charge,
the charge of the background axion field and $n$ magnetic charges,
respectively.  When the NUT parameter is set to zero, a rotating dyonic black
hole solution endowed with a rotating axion--dilaton field arises.

\section{Conclusions and Discussion}
In this letter we have obtained a class of stationary extremal
solutions that generalize the IWP class of EM theory for the
$d+3$--dimensional heterotic string compactified on a $d$--torus using
the MEP formalism. The physical charges of the field system saturate the BPS
bound as a consequence of the extremal character of the found solutions.
These solutions are expressed in terms of $2(d+n)$ ($n\geq d+1$ being
the number of Abelian vector fields) independent real parameters
related to physical charges of the field system.

In a special case the found solutions correspond to a dyonic asymptotically
Taub--NUT rotating field configuration with axial symmetry. This object is
formed by the following rotating fields: the gravitational and Kalb--Ramond
fields, the axion/dilaton, as well as by $(d+1)\times n$ electromagnetic
charges.

Among these solutions we identify (by requiring the asymptotic flatness
condition to be satisfied) a class of rotating dyonic black hole--type
configurations. All the four--dimensional rotating black hole solutions
of this type develope naked singularities before the BPS bound is reached.
So there is no horizon hidding the singularity. This fact makes impossible
the study of thermal properties of extreme rotating objects since their
entropy is proportional to the horizon area.
If indeed, the rotational parameter vanishes, the solutions become
static. The thermodynamical properties of these objects are well--known (see
\cite{y}
, for instance).

In principle, the MEP formalism allows one to extract a richer class of
solutions by considering a more complete ansatz (with more general complex
$M$ and $b$). In this case, the Kaluza--Klein
and antisymmetric fields will have different charges, for example.
Moreover, the number of independent parameters will increase since we will
have more components of $M$ and more independent electromagnetic charges.
On the other hand, one can consider multi--center solutions or linear
combinations of Legendre polinomials as solutions of the Laplace equation.

At the end of the letter we would like to stress one special property of our
solutions. One can see that the electric potentials $A_t^I$ from Eqs. (22)
have no dipole momentum in contrast with the components of the vector
potentials $A_{\vp}^{(D)I}$. This is not an intrinsic property of
the equations of motion. Actually, it is well known that the equations of
motion posses the $SL(2,{\bf R})$ symmetry that interchanges the electric
and magnetic sectors. In this work we show how in the framework of the
constructed solutions one can interprete the magnetic dipole momenta as a
result of the rotation of the electric charges (a similar situation takes
place for the axion field, where its dipole momentum is defined by the
rotating dilaton charge). To achive this symmetry in the solution one must
choose a more general solution of the Laplace equation (10). It seems that
taking into account the next dipole term in the solution (10) it is possible
to recover the lost symmetry. However this requires further investigation.

\section*{Acknowledgments}
We would like to thank our colleagues of NPI and JINR for
encouraging us during the performance of this letter.
A.H. was partially supported by CONACYT and SEP.


\end{document}